\documentstyle[12pt,psfig,epsf]{article}
\newcommand{\nn}{\nonumber}
\newcommand{\be}{\begin{equation}}
\newcommand{\ee}{\end{equation}}
\newcommand{\bea}{\begin{eqnarray}}
\newcommand{\eea}{\end{eqnarray}}

\begin{document}
\newcommand{\nd}[1]{/\hspace{-0.5em} #1}
\begin{titlepage}

\begin{flushright}
NTUA 69/98 \\
OUTP-98-10P \\
hep-lat/9802037 \\
\end{flushright}

\begin{centering}
\vspace{.1in}
{\Large {\bf Dynamical Flavour Symmetry Breaking by a Magnetic 
Field in Lattice QED$_3$ \\}}
 
\vspace{.2in}
{\bf  K. Farakos} and {\bf G. Koutsoumbas}
 
\vspace{.05in}

National Technical University of Athens, 
Physics Department,
Zografou Campus
GR-157 80, Athens, Greece, \\

\vspace{.05in}
and \\
\vspace{.05in}
{\bf N.E. Mavromatos}$^{*}$ \\
\vspace{.05in}
University of Oxford, Department of (Theoretical) Physics, 
1 Keble Road OX1 3NP, Oxford, U.K. \\

\vspace{.2in}
{\bf Abstract} \\
\vspace{.05in}
\end{centering}
{\small We perform a lattice study, in the quenched approximation,
of dynamical mass generation in a system of relativistic (Dirac) fermions,
coupled to an Abelian gauge field in ($2+1$)-dimensions, in the presence of 
an external (constant) magnetic field, perpendicular to the spatial planes. 
It is shown that a strong magnetic field catalyzes chiral symmetry breaking,
in agreement with results in the continuum. 
The r\^ole of the higher-Landau poles in inducing a critical temperature
above which the phenomenon disappears is pointed out. 
We also discuss the implications of this model on the opening of a gap 
in doped antiferromagnetic superconductors. }
\vspace{.3in}

\begin{flushleft} 
February 1998 \\
$^{(*)}$~P.P.A.R.C. Advanced Fellow. \\
\end{flushleft} 

\end{titlepage}

\section{Introduction}
In $(2+1)$-dimensional gauge theories with relativistic fermions
chiral symmetry can be defined only in the case of an 
{\it even} number of {\it massless} fermion species, $2N$,  
in the so-called $4 \times 4$
{\it reducible} Dirac formalism~\cite{app}.
Fermion mass terms break the chiral symmetry 
$U(2N) \rightarrow U(N)\times U(N)$. 
To preserve parity one needs to have masses with opposite sign 
between the fermion species. This is the only type of mass terms that 
is allowed to be generated {\it dynamically} in {\it vector-like theories},
as a result of energetics arguments~\cite{Vafa}. 
 
The problem of chiral symmetry breaking in $(2+1)$-dimensional 
{\it continuum} QED$_3$ in the presence of {\it external magnetic fields} 
has been 
studied extensively in the recent literature~\cite{miransky,shpagin},
following analyses in four dimensional gauge 
theories~\cite{chodos,miransky,fourdim}.
According to these findings, a strong enough magnetic field
acts as a strong catalyst of chiral 
symmetry breaking, which occurs even in the limit
of weak gauge couplings. We 
remind the reader that, in the absence of such strong fields,
dynamical chiral symmetry breaking (i.e. fermion mass generation) occurs only 
in {\it supercritical theories}~\cite{app}, i.e. for 
dimensionless gauge couplings 
that are stronger than some critical value $g_c$~\footnote{In $(2+1)$-dimensions the effective dimensionless gauge coupling is actually 
the inverse of the fermion flavour number, $1/N$. Thus supercriticality 
in the coupling means actually the existence of a critical number 
of flavours $N_c$, below which dynamical chiral symmetry breaking is possible.
Continuum Schwinger-Dyson analyses~\cite{app}, 
in the large $N$ approximation,  
have indicated that $N_c \simeq 32/\pi^2$. This result has also been 
argued to be correct on the basis of lattice analyses~\cite{kocic}.}.
The effect of the strong magnetic field is, according to the analysis 
of \cite{miransky,fourdim}, to reduce $g_c$ to zero. This phenomenon was 
termed in \cite{miransky} {\it magnetic catalysis}. 

This phenomenon may be important in the physics of high-temperature superconductivity, as pointed out in ref. \cite{fmtb}. Indeed, it is believed that 
high-temperature superconductors may have some strong connection with 
relativistic $(2+1)$-dimensional gauge theories~\cite{dorey,KR,fm,fmtb}.
This belief is enforced recently, as  a result of experiments indicating that
the high-temperature superconductivity is actually due to the opening of 
a $d$-wave  gap in the hole (fermion) spectrum~\cite{dwave}. As is well known, $d$-wave
superconductivity is characterized by lines of {\it nodes} of the 
gap. In the gauge theory approach 
to doped antiferromagnets, such 
hole (charged) excitations about such nodes are 
described by Dirac fermions~\cite{dorey,fm}, coupled to gauge fields
that describe the magnetic interactions among the charged degrees of freedom,
in a spin-charge separating formalism~\cite{Anderson}. 

The models of ref. \cite{dorey,fmtb} 
describe the dynamical opening of a fermion gap at those nodes. 
Moreover, the theoretical analysis of \cite{fmtb} indicated that
the dynamical fermion gap, due to the statistical gauge interactions,
is enhanced in the presence of strong external magnetic fields,
as expected from the general theory of \cite{fourdim,miransky,shpagin}.  
The analysis of \cite{fmtb} also indicated the existence of a 
critical temperature above which chiral symmetry is restored
in the model. This phase transition occurs well within the 
superconducting phase of the model and pertains only to the 
fermion gap at the nodes of the $d$-wave gap~\footnote{We note
that the problem of looking at charged femions in the presence of an 
external field is compatible with the superconducting nature of 
the high-temperature superconductors. These materials are 
known to be strongly type II, which implies that the 
Meissner effect penetration depth in the superconducting phase
is quite large, a few thousands of Angstr\"oms.}. 
For the model of \cite{fmtb},
in the absence of external magnetic fields, the fermion gap at the nodes
disappears at temperatures $0.1$~K at most; in contrast, under the 
influence of a strong magnetic field $1~{\rm Tesla} < B < 20~{\rm Tesla}$,
the fermion gap at the nodes remains up to temperatures of order $30$ K. 

It is worth pointing out that 
recent experiments have indicated the presence of such gaps
at the nodes of a conventional $d$-wave gap~\cite{ong}.
Such experiments involved magnetic fields of the above order of magnitude,
and they showed clearly the existence of a gap up to temperatures
that are in close agreement with the above theoretical results. 
Although there may be alternative explanations for the phenomenon, 
the above qualitative agreement between theory and experiment
prompts further studies of the magnetic catalysis phenomenon.

In addition to the above-mentioned motivation from Condensed Matter 
Physics, the study of the magnetic catalysis phenomenon in
three-dimensional gauge theories is also useful from the 
point of view of high-temperature four-dimensional gauge 
theories in the Early Universe, 
or the physics of high-temperature quark-gluon plasma,
in the presence of external fields, since both cases
may be effectively described by three-dimensional gauge models. 

Whatever the motivation might be, it becomes clear that the 
problem of the formation of a fermion condensate in the presence of strong
external fields is quite relevant for physical applications. 
So far, as far as we are aware of, 
the analysis was restricted to the continuum 
formalism~\cite{miransky,shpagin}. It is the purpose of this work 
to present a first, preliminary, analysis of the phenomenon on the lattice,
in the quenched approximation in the case of 
$(2+1)$-dimensional $QED$. 

We shall discuss here the formation of parity-invariant 
fermion condensates. 
In our analysis we shall trigger the 
phenomenon of chiral symmetry breaking by adding 
bare masses which preserve parity,
and then extrapolate the results to the zero bare mass case.  
In this way we find only 
magnetically-induced condensates which preserve parity.  
Notice that in the presence of an external magnetic field,
the theory ceases to be vector-like, and hence the theorem of 
\cite{Vafa} no longer applies, and one could in principle 
have parity-violating fermion condensates. This phenomenon
was conjectured by Laughlin~\cite{Laughlin}  
as happening in the case of the high-temperature 
superconductors, offering a possible explanation of the 
results of the 
experiments of \cite{ong}. 
Our results in this work, though,
indicate that 
parity-violating induced condensates might not be necessary 
for a possible explanation of the situation in the experiments
of \cite{ong}.

\section{Results in the continuum from a novel perspective}

In this section we describe briefly the theoretical
formalism in the continuum~\cite{miransky,shpagin}. One follows 
essentially the method of Schwinger~\cite{schwing}, 
by looking at the coincidence limit of the fermion propagator (in configuration space), ${\rm Lim}_{x \rightarrow y}~S(x,y)|_B$,
in the presence of a constant external field, $B$.
We start first from the free-fermion case, i.e. the case where the
fermions interact only with the external constant field. 
The external gauge potential is then given by: 
\be
A_\mu = -Bx_2\delta_{\mu1}
\label{extpot}
\ee 
and the Lagrangian is:
\be
  L=\frac{1}{2}{\overline \Psi}(i \gamma^\mu (\partial_\mu - ieA_\mu^{ext}) - m)\Psi  
\label{lagrangian}
\ee
where $m$ is a {\it parity conserving } bare fermion 
mass, and the $\gamma$-matrices
belong to the reducible  $4\times 4$ representation, appropriate 
for an even number of fermion species formalism~\cite{app,dorey,fm}. 

The induced fermion condensate is given by the coincidence 
limit of the fermion propagator~\footnote {We define for convenience the
condensate with a plus sign, rather than the usual minus sign.}
\bea 
 &~&  <0|{\overline \Psi} \Psi |0>|_B = + Lim_{x \rightarrow y}trS(x,y)|_B \nn \\
 &~& {\rm where}~~ S(x,y)=<0|T{\overline \Psi (x)}\Psi (y)|0>|_B
 \label{propagator}
\eea

Following the proper time formalism of Schwinger~\cite{schwing},
the propagator $S(x,y)|_B$ 
in the presence of a constant external magnetic field 
can be calculated exactly~\cite{miransky}:  
\bea
&~&S(x,y)|_B = {\rm exp}\left(ie\int_x^y A^{ext}_\mu dz^\mu \right)
{\tilde S}(x-y)|_B \nn \\
&~&{\tilde S}(x)|_B =\int_0^\infty \frac{ds}{8(\pi s)^{3/2}}{\rm exp}[-i
\left(\frac{\pi}{4} + sm^2 \right)]\times {\rm exp}[-\frac{i}{4s}
x_\nu C^{\nu\mu}x_\mu] \nn \\
&~&\times [\left(m + \frac{1}{2s}\gamma^\mu C_{\mu\nu}x^\nu-\frac{e}{2}\gamma^\mu
F_{\mu\nu}^{ext}x^\nu \right) \times \left(esB{\rm cot}(eBs)-\frac{es}{2}
\gamma^\mu\gamma^\nu F_{\mu\nu}^{ext} \right)]
\label{proptime}
\eea
where $F_{\mu\nu}^{ext}$ is the Maxwell tensor corresponding to the 
external background gauge potential (\ref{extpot}), and 
$C_{\mu\nu}=\eta^{\mu\nu}+ ([F^{ext}]^2)^{\mu\nu}[1-eBs{\rm cot}(eBs)]/B^2$,
and the line integral is calculated  along a straight line. 
A useful expression, to be used in the following, is the Fourier 
transform ${\tilde S}_E(k)|_B$ in Euclidean space 
of ${\tilde S}(x-y)|_B$ ~\cite{miransky}:
\bea
&~&{\tilde S}_E (k)|_B =-i\int_0^\infty ds {\rm exp}[-s(m^2 +k_0^2 + k_\perp^2
\frac{tanh(eBs)}{eBs}] \nn \\
&~&\times \{[-k_\mu\gamma^\mu + m -i(k_2\gamma_1 -
k_1\gamma_2)tanh(eBs)]\times [1 -i\gamma_1\gamma_2tanh(eBs)]\}
\label{eucls}
\eea
We define 
\be
\Delta <{\overline \Psi} \Psi>|_B \equiv
<0|{\overline \Psi} \Psi |0>|_B-<0|{\overline \Psi} \Psi |0>|_{B=0}
\ee
A straightforward calculation, then, yields in (2+1)-dimensional
space-time the following result for
$\Delta <{\overline \Psi} \Psi>|_B$ ~\cite{miransky}:
\bea
&~&\Delta <{\overline \Psi} \Psi>|_B
=\frac{i}{(2\pi)^3}\int d^3k tr [{\tilde S}_E(k)|_B-{\tilde S}_E(k)|_{B=0}]
\nn \\
&~&=\frac{4m}{(2\pi)^3}
\int d^3k \int_0^\infty ds [ {\rm exp}\{-s(m^2 + k_0^2 +k_\perp^2 
\frac{tanh(eBs)}{eBs})\}-{\rm exp}\{-s(m^2 + k_0^2 +k_\perp^2)\} ] 
\nn \\
&~&=\frac{m e B}{2 \pi^{\frac{3}{2}}}
\int_0^\infty ds \frac{e^{-s m^2}}{\sqrt{s}}
[coth(e B s) -\frac{1}{e B s}]
\label{condensate}
\eea

In the formulae above we have used the notations: $k_\perp^2=k_1^2+k_2^2.$
The limit of the final expression as $m \rightarrow 0$ is easily found to be
$\frac{e B}{2 \pi}.$ We note here that the infinities in 
$<{\overline \Psi} \Psi>$ are already present for zero magnetic field, so by
performing the subtraction to form $\Delta <{\overline \Psi} \Psi>|_B$ we 
get an automatically finite result and there is no need to introduce explicit 
cut-offs in the integrals. 

We show, in figure 1, the outcome of the numerical calculation of 
(\ref{condensate}) for three different masses. The magnetic field in the
following figures will be represented by the ratio 
$b \equiv \frac{e B}{\pi}.$ \footnote{The denominator 
$\pi$ has been included here to facilitate the comparison with 
the lattice results: there the quantity $b$ is defined as the ratio
$\frac{eB}{eB|_{max}},$ where $eB|_{max}$ is the maximum  
magnetic field (in lattice units) that can be achieved on the lattice; 
this maximal value is $\pi.$}
There is a striking result, 
depicted in this figure, namely that $\Delta <{\overline \Psi} \Psi>|_B$ 
{\it increases} for decreasing mass. This may be understood through 
the remark that the response of the system to an external field is 
bigger for small masses. An important point is that the value of this 
quantity has a finite limit as $m \rightarrow 0,$ the value 
$\frac{e B}{2 \pi},$ referred to above.

%%%%%%%%%%%%%%%%%%%%%%%%%%%%%%%%%%%%%%%%%%%%%%%%%%%%%%%%%%%%%%%%
\begin{figure}
\centerline{\hbox{\psfig{figure=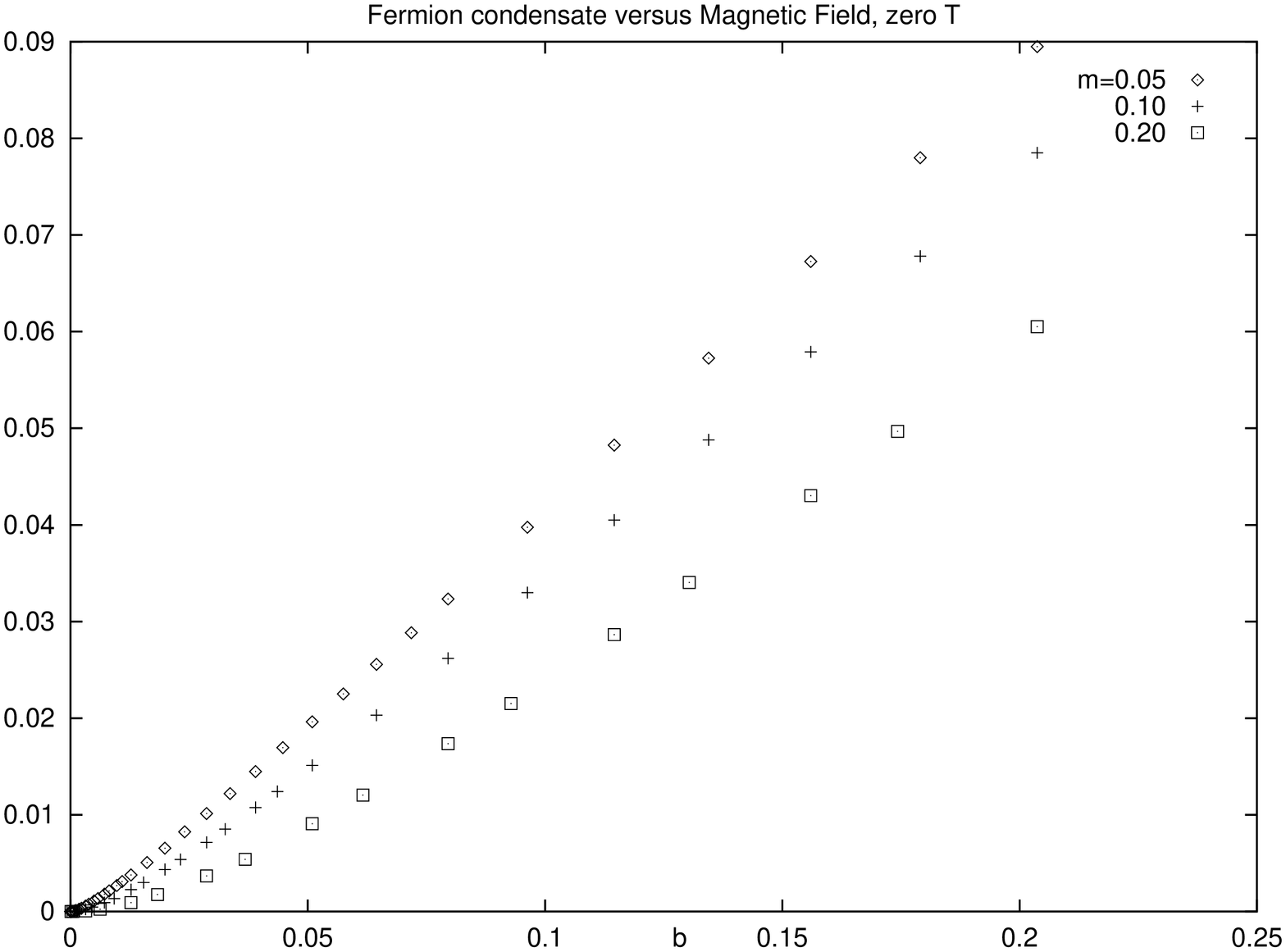,height=9cm,angle=-90}}}
\caption[f1]{$<{\overline \Psi} \Psi>$ versus $\frac{e B}{\pi}$ 
at zero temperature}
\label{f1}
\end{figure}
%%%%%%%%%%%%%%%%%%%%%%%%%%%%%%%%%%%%%%%%%%%%%%%%%%%%%%%%%%%%%%%%

Further insight in the model, as well as the tools to investigate the 
finite temperature issue, is offered by the expansion of 
the propagators in terms of the Landau level contributions. 
The relevant expansion reads:

\be
 {\tilde S}_E(k)|_B=-i e^{\frac{-k_\perp^2}{e B}} 
\sum_{n=0}^{\infty} (-1)^n \frac{D_n(m,B,k)}{k_0^2+M_n^2(B)},
\ee
with $ M_n^2(B) \equiv m^2+ 2 e B n$ and
$ Tr D_n(m,B,k)=4 m [L_n(\frac{2 k_\perp^2}{e B})
-L_{n-1}(\frac{2 k_\perp^2}{e B})]. $

We need the
integrals over the momenta $k_1,~k_2,$ which are calculated using 
the formula:

\be
 \int dk_1 dk_2 e^{-\frac{k_\perp^2}{e B}}[L_n(\frac{2 k_\perp^2}{e B})
-L_{n-1}(\frac{2 k_\perp^2}{e B})] = (-1)^n 2 \pi e B.
\ee

What is left is the sum over the $k_0$ momentum. We consider the 
case of zero temperature (the case we have just treated), which means 
one should perform the integral over $k_0.$ The result reads:

\be
<{\overline \Psi} \Psi>|_B 
=\frac{e B}{2 \pi} 
+\frac{e m B}{\pi} \sum_{n=1}^\infty \frac{1}{M_n(B)} =
-\frac{eB}{2\pi}+\frac{m\sqrt{2eB}}{2\pi}\zeta (\frac{1}{2}, \frac{m^2}{2eB})
\label{zeta} 
\ee
where $\zeta (z, q)$ is one of the Riemann $\zeta$ functions,
which is understood to be defined through appropriate analytic 
continuation~\cite{ryzhik}. 
The first term in the middle 
part of the equality (\ref{zeta}) 
is mass independent and coincides with the previous 
result in the massless limit, while the remaining part of the 
expression vanishes in this limit. Thus, for the zero-temperature 
case, we find that in three dimensional gauge theories 
a strong magnetic field (such that $|eB| $ is much larger 
than any other mass scale in the problem), may induce a fermion 
gap~\footnote{Notice an important difference from the 
corresponding four-dimensional 
problem, where the corresponding fermion condensate reads: 
\be 
  <{\overline \Psi}\Psi > 
\sim -|eB|\frac{m}{4\pi^2} (
{\rm ln}\frac{\Lambda ^2}{m^2} + {\cal O}(m^0) ) \rightarrow 0
\quad ; \quad m \rightarrow 0
\label{fourdim}
\ee
and tends to zero as the bare mass $m \rightarrow 0$ , 
for a fixed (ultraviolet ) scale $\Lambda$. However,
the quantum dynamics of the electromagnetic field may drastically 
change the situation if treated non-perturbatively~\cite{miransky,fourdim}.}. 

It is worthy of examining some physically interesting limiting cases 
of (\ref{zeta}), keeping however $eB$ finite. 
First, consider the case of weak magnetic fields
(compared to the (bare) fermion mass $m$): $eB << m^2$. I that case 
one may expand $M_n^{-1} \simeq m^{-1}\left(1 - \frac{eB}{m^2}n + \dots \right)$. The resulting sums can be regularized by means of the 
Riemann's second $\zeta$ function, $\zeta (z)=\sum _{n=1}^{\infty}n^{-z}$,
using~\cite{ryzhik}: $\zeta (0)=-\frac{1}{2}$, $\zeta (-1)=-1/12$
(both being understood 
by means of appropriate analytic continuation). The result is:
\be
 \Delta <{\overline \Psi} \Psi>|_{eB << m^2} \simeq  \frac{(eB)^2}{12m^2\pi}
+ \dots 
\label{smallB}
\ee
with the $\dots $ indicating subleading terms of higher order in $eB/m^2$. 
Thus, for weak magnetic fields the induced gap scales
quadratically with the external field intensity, for massive fermions.

In a similar manner, for strong 
magnetic fields $eB >> m^2$, one finds 
\be
\Delta <{\overline \Psi} \Psi>|_{eB >> m^2} \simeq  \frac{(eB)}{2\pi}
\left(1 + \zeta (\frac{1}{2}) \frac{\sqrt{2}m}{\sqrt{eB}} - \frac{\sqrt{2}m^3}{[eB]^{3/2}}\zeta (\frac{3}{2}) 
+ \dots \right) 
\label{largeB}
\ee
where $\zeta (\frac{1}{2}) \simeq -1.46$, $\zeta (\frac{3}{2}) \simeq +2.61.$. 
Notice that despite the fact that the various terms in 
(\ref{largeB}) come with alternating signs, however 
{\it there is no critical value} of the field, consistent with 
the approximation $m/\sqrt{eB} << 1,$ 
at which the condensate in (\ref{largeB}) {\it changes sign}. 

We now come to  the model at finite temperature. The integral 
over $k_0$ should be replaced by a sum over $k_0 = 2 \pi (n+\frac{1}{2}) T.$
The $T=0$ result is modified to:
\be
<{\overline \Psi} \Psi>|_T= \frac{e B}{2 \pi} tanh(\frac{m}{2 T}) 
+\frac{e m B}{\pi} \sum_{n=1}^\infty \frac{1}{M_n(B)} tanh(\frac{M_n(B)}{2 T}),
\label{finiteT}
\ee
thereby implying the {\it absence} of the condensate for 
{\it any finite temperature} if 
the bare infrared cut-off mass $m,$ for finite constant $B.$

In a similar fashion, as for the zero-temperature case, one may perform 
the summation over the Landau poles for certain limiting cases.
The physically interesting case is the one for which one assumes
a strong field $eB >> m^2$ and a temperature regime $T >> m$, but 
such that $eB >> T^2$. In such a case $tanh (M_n(B)/2T) \simeq 1$ for all $n$, 
whilst $M_n (B) \simeq \sqrt{2eBn}$, to leading order. 
Hence, 
\be
\Delta <{\overline \Psi} \Psi>|_T \simeq 
\frac{e B}{4 \pi} \frac{m}{T} 
-\frac{1.46m\sqrt{eB}}{\sqrt{2}\pi} 
\ee
using $\zeta (1/2)=-1.46$. Notice the existence of a
{\it critical temperature}, for non-zero $m$, 
\be 
      T_c \simeq \frac{1}{4}\sqrt{eB}
\label{critical}
\ee
above which the condensate vanishes~\footnote{To be precise,
at 
$T=T_c$ the condensate changes sign. However, given that 
in three-dimensional gauge 
theories $<{\overline \Psi}\Psi > \propto E^2$, with $E$ the 
(magnetically-enhanced) fermion energy gap, such a change in sign is 
interpreted as implying 
that 
above $T_c$ thermal fluctuations dominate, and, hence, prevent 
chiral symmetry breaking. It is in this sense that we used above the 
terminology `critical temperature'.}. 
The order of magnitude of the temperature
is consistent with the approximations made in the derivation of 
(\ref{critical}), which suggests an important r\^ole for the higher Landau
levels at finite temperatures, 
\footnote{Especially as far as 
the superconductivity phenomenon is concerned, there is a natural 
upper bound for the magnetic field, $B_c,$ above which 
superconductivity is destroyed. For realistic cases, 
the critical field in the high-temperature
superconducting oxides is of order of $20$ Tesla.
For such fields the inclusion of the higher Landau 
poles seems necessary.}
namely that they {\it prevent}
dynamical mass generation above a certain temperature. This is consistent 
with the fact that for $B \rightarrow \infty$, where only the lowest Landau 
poles are retained, there is dynamical mass generation for all $T < \infty$. 
In the context of a possible application of this phenomenon 
to high-temperature superconductors~\cite{fmtb}, we note that $\sqrt{eB}$ 
scaling of a critical temperature with the external field intensity 
are reported in the experiments of \cite{ong}. Notice however, that 
the existence of extra interactions, that may lead to dynamical mass 
generation for fermions, as in the $SU(2)\times U_S(1)$ 
model of \cite{fm}, the scaling 
of the critical temperature with $B$ may be different~\cite{fmtb}.  

We now proceed with a different regularization, based on the remark that all 
infinities in the model lie in the $T=0$ sector, so that the difference 
\be
<{\overline \Psi} \Psi>|_T-(<{\overline \Psi} \Psi>|_{T=0}) 
=\frac{e B}{\pi} \frac{1}{1+e^{\frac{m}{T}}}
+\frac{2 e m B}{\pi} \sum_{n=1}^\infty \frac{1}{M_n(B)} 
\frac{1}{1+e^{\frac{M_n(B)}{T}}} ,
\label{series}
\ee
of the condensate at finite temperature, as given by (\ref{finiteT}), and its 
zero temperature limit should be finite. On the other hand, the 
zero temperature part has already been regularized above by the subtraction 
of the $B=0$ part and the finite result is given in equation 
(\ref{condensate}). Thus, we may calculate $\Delta<{\overline \Psi} \Psi>|_T$
substituting for $\Delta<{\overline \Psi} \Psi>|_{T=0}$  
the regularized expression of (\ref{condensate}) and for the 
finite difference the series of (\ref{series}).
The result of these calculations is depicted in figure 2. 

%%%%%%%%%%%%%%%%%%%%%%%%%%%%%%%%%%%%%%%%%%%%%%%%%%%%%%%%%%%%%%%%
\begin{figure}
\centerline{\hbox{\psfig{figure=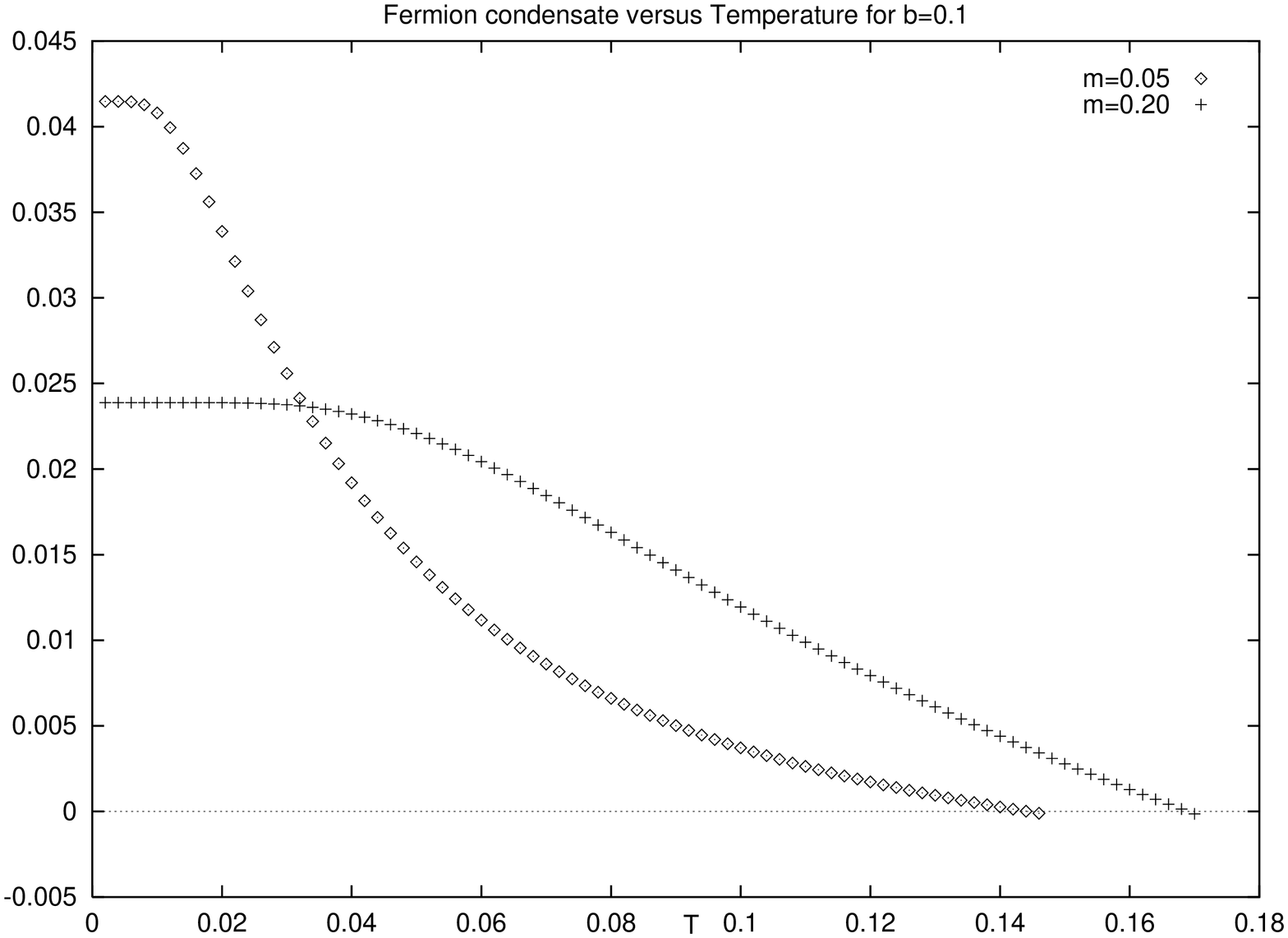,height=9cm,angle=-90}}}
\caption[f2]{$<{\overline \Psi} \Psi>$ versus 
temperature at fixed magnetic field}
\label{f2}
\end{figure}
%%%%%%%%%%%%%%%%%%%%%%%%%%%%%%%%%%%%%%%%%%%%%%%%%%%%%%%%%%%%%%%%

We have used two
masses, $m=0.20$ and $m=0.05.$ We remark that the regularization scheme we 
have adopted seems to work quite well. 
We observe that (say for the larger mass, where
the effects are more transparent) one may distinguish three 
temperature regimes, delimited by $\tau_1 \equiv T a \simeq 0.04$ and 
$\tau_2 \equiv T a \simeq 0.17.$
Below $\tau_1$ the condensate is the same as for zero temperature, thus if 
one uses lattices of time extent equal roughly to $\frac{1}{\tau_1},$
which equals about 25 in the $m=0.20$ case, one should get zero
temperature results. After that the condensate starts decreasing 
and vanishes at $\tau_2.$ Thus, for lattices with temporal extent of
about $\frac{1}{\tau_2}$ (about 6 in the $m=0.20$ case) one should 
expect the vanishing of the condensate. For the smaller value of the mass, 
0.05, the zero temperature value is larger, in agreement with the
results in figure 1, but it decreases more steeply. The corresponding
values for $\tau_1$ and $\tau_2$ are 0.01 and 0.145 respectively, 
yielding the corresponding temporal extents 100 and 7. Comparing the results
for the two masses we observe that for small temperatures (corresponding
to large enough lattices) the condensate for the smallest masses are the 
largest, as explained in figure 1; however, if the lattice is relatively
small, corresponding to a high temperature, it is expected that the 
condensates for the ``big" masses will take over. This could be used 
as a check that the lattice size used is large enough.

We have plotted the value of the critical temperature $T_c$ (denoted by 
$\tau_2$ above) versus the magnetic field in figure 3.

%%%%%%%%%%%%%%%%%%%%%%%%%%%%%%%%%%%%%%%%%%%%%%%%%%%%%%%%%%%%%%%%
\begin{figure}
\centerline{\hbox{\psfig{figure=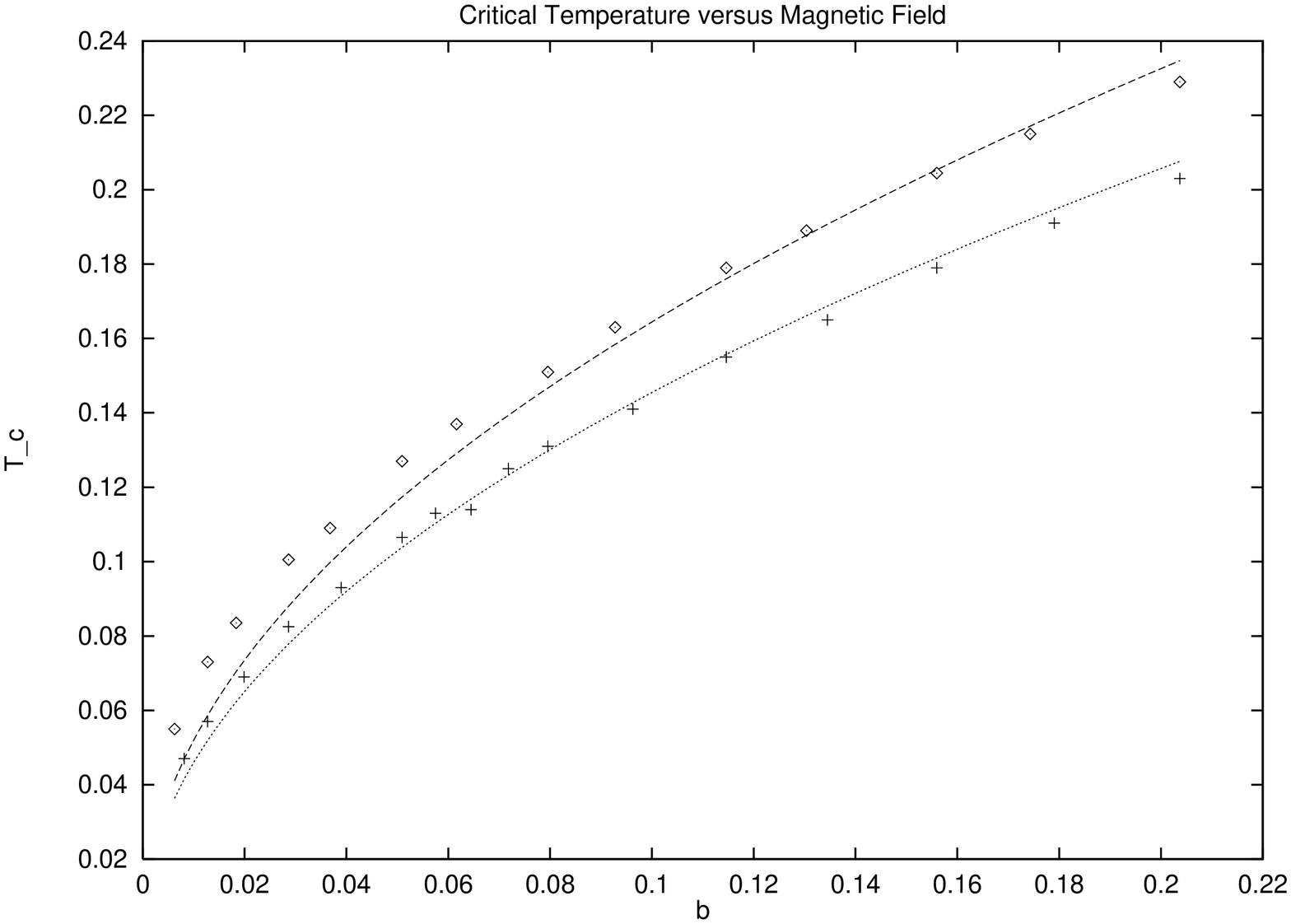,height=9cm,angle=-90}}}
\caption[f3]{Critical temperature versus 
magnetic field for two masses: the upper data and fit correspond to 
$m=0.20,$ the lower data and fit to $m=0.05.$}
\label{f3}
\end{figure}
%%%%%%%%%%%%%%%%%%%%%%%%%%%%%%%%%%%%%%%%%%%%%%%%%%%%%%%%%%%%%%%%

We have used data for two masses and performed a square root fit: 
$T_c= \kappa(m) \sqrt{b}=\frac{\kappa(m)}{\sqrt{\pi}} \sqrt{(e B)}.$ 
It turns out that $\kappa(0.05)=0.46$ and
$\kappa(0.20)=0.52,$ so that $\frac{\kappa(0.05)}{\sqrt{\pi}} \simeq 0.26,$ 
$\frac{\kappa(0.20)}{\sqrt{\pi}} \simeq 0.29.$
We observe that the fit is better for the small 
mass, which is quite reasonable, since the square root behaviour has 
been found in the limit of big magnetic field as compared to the mass squared.
The big values of the magnetic field have been fitted, so the discrepancies
occuring in for the bigger mass take place at small magnetic fields.
The proportionality constants 0.26 and 0.29 given by this approach is
equal to a very good approximation with the value $\frac{1}{4}$ 
yielded by the $\zeta$
function regularization (\ref{critical}), applicable when the magnetic 
field is strong. We note that strong field means value of the ratio
$\frac{eB}{m^2}$ much greater than 1. In the case we are considering,
the value $b=0.1$ corresponds to $\frac{e B}{m^2} \simeq 8,$ for m=0.20 and 
to $\frac{e B}{m^2} \simeq 128,$ for m=0.05.
Thus the two different approaches 
agree quite well and reproduce the magnetic field dependence found 
experimentally.

\section{The Lattice Model}

We now describe briefly the field-theoretical model associated to
the problem of superconductivity as formulated above. The relevant
three-dimensional Lagrangian is:
\be
L=-\frac{1}{4} (f^S_{\mu \nu})^2 +{\overline \Psi} D_\mu \gamma_\mu \Psi -m
{\overline \Psi} \Psi,
\ee
with $D_\mu$  the covariant derivative. The field strength corresponds
to the ``statistical" gauge field $\alpha_\mu^S.$ The mass term has been
introduced by hand to facilitate the lattice treatment and the limit $m
\rightarrow 0$ will be considered in the end. It has already been explained 
that the
Clifford algebra is reducible in three dimensions and the model may also
be represented by $2 \times 2$ matrices acting on two species of fermions
with opposite mass parameters. The generators ${\bf 1},$ $ \gamma_3,$
$\gamma_5$ and $\Delta \equiv i \gamma_3 \gamma_5$ generate a global $SU(2)
\times U(1)$ symmetry of the Lagrangian (if m is set to zero). We observe
that one may write down two kinds of mass term: the parity conserving mass
term $m_C {\overline \Psi} \Psi$ and the parity violating one: $m_V
{\overline \Psi} \Delta \Psi.$ It is easy to see that the parity conserving
mass term breaks the global $SU(2)$ symmetry, while the parity violating
mass is an $SU(2)$ singlet. The $U(1)$ field acting on the fermions may
consist of two parts, the electromagnetic $U_E(1),$ generated by an
external electromagnetic potential $A_\mu,$ and the aforementioned statistical
$U_S(1),$ associated with $\alpha_\mu^S,$ responsible for the 
fractional statistics of the holon
excitations of the doped antiferromagnet.

The model with an external
homogeneous magnetic field has been examined in 
\cite{miransky} in the two-component
formalism for the fermions; 
it was found that the (parity violating) magnetic field gives
rise to parity violating mass terms with opposite signs for the two
fermion species. Thus, in the four-component formalism, the dynamically
generated parity violating mass terms are expected to sum up to zero. On
the other hand, the effect of mass generation will manifest itself in an
enhancement of the parity conserving mass, as compared to the free fermion 
field case.

We would like to study the above effects by lattice methods; this will
free us from the need of the approximations used in 
\cite{miransky} and will permit us
to include the statistical gauge
field, on top of the external electromagnetic field. The lattice action,
corresponding to the continuum Lagrangian, may be written as:
\bea
&~&S =\beta_G \sum_{P} (1-Tr U_P) + \sum_{n,n^\prime} {\overline \Psi}_n 
Q_{n,n^\prime} \Psi_{n^\prime}
\nn \\
&~&Q_{n,n^\prime} = \delta_{n,n^\prime}-K \sum_{\hat \mu} 
[\delta_{n^\prime,n+\hat \mu} (r+\gamma_{\hat \mu}) U_{n {\hat \mu}} 
V_{n {\hat \mu}} + \delta_{n^\prime,n-\hat
\mu} (r-\gamma_{\hat \mu}) U_{n-\hat \mu, \hat \mu}^\dagger 
V_{n-\hat \mu, \hat \mu}^\dagger].
\label{lact}
\eea
The indices $n,~n^\prime$ are triples of integers, such as $(n_1,~n_2,~n_3),$ 
labeling the lattice sites, while $\mu$ denotes directions.
$r$ is the Wilson parameter, $K$ the hopping parameter, 
$U_{n {\hat \mu}} \equiv
e^{i g a \alpha^S_{n {\hat \mu}}},~~V_{n {\hat \mu}} 
\equiv e^{i e a A_{n {\hat \mu}}},$ with
$\alpha^S_{n {\hat \mu}}$ the statistical gauge 
potential and $A_{n {\hat \mu}}$ the
external electromagnetic potential. $U_P$ is the product of the links 
surrounding a plaquette, as usual, and $\beta_G$ is related to the  three
dimensional $U_S(1)$ coupling constant by $\beta_G=\frac{1}{g^2 a},$ while
we denoted by $e$ the four dimensional (dimensionless) electromagnetic 
coupling of $U_E(1).$ In our treatment we will use naive
Wilson fermions, so we set $r=0.$ 

On the other hand, we would like to impose an external homogeneous 
magnetic field in the (missing) $x_3$ direction, thus we
choose the external gauge potential in such a way that the plaquettes in
the $x_1 x_2$ plane equal B, while all other plaquettes equal zero. This
may be achieved in several ways, related with each other by gauge
transformations. The value of B is restricted, due to the fact that the
flux through the entire $x_1x_2$ section of the lattice should equal an 
integer multiple of $2 \pi,$ so
for the $x_1 x_2$ plane with area $N^2$ we have: $B=\frac{2
\pi}{N^2} l,$ with $l$ some integer between 0 and $\frac{N^2}{2}.$
(The model with integers $l$ between $\frac{N^2}{2}$ and $N^2$ is 
equivalent to the model with integers $N^2-l$ which fall into 
the previous category.) We remark here that the maximum value the magnetic
filed $B$ can take is $\pi.$ The physical magnetic field $B_{phys}$ is
related to $B$ through $B=e \alpha^2 B_{phys},$ and the physical field may
go to infinity, as it should, if the lattice spacing $\alpha$ goes to zero,
while $B$ is kept constant.

However, the above considerations are complicated by the fact that 
the lattice is a torus, that is a closed surface, and the Maxwell equation 
$div {\bf B} =0$ implies that the magnetic flux through the lattice 
should vanish. This means that, if periodic boundary conditions are
used for the gauge field,\footnote {An alternative way out might be, 
of course, the use of twisted boundary conditions.} the gauge field 
configuration will have zero total flux, so the (positive, say) flux 
$B,$ penetrating 
the majority of the plaquettes, will be accompanied by a compensating 
negative flux in a small number of plaquettes. This point will become 
clear in a while. 

To begin with, we consider a configuration giving flux B in each plaquette. 
In analogy with the continuum, one would consider something like:
$a e A_1 = -B n_2,~~A_2 = 0, ~~A_3=0,$ with $B=\frac{2 \pi}{N^2} l.$
However, due to the periodic boundary conditions, the plaquettes 
in the $x_1x_2$ plane starting at the point $(n_1,n_2)$ with $n_2 = N,$ 
will not have flux equal to B. To cure this, one modifies a little bit 
the configuration by demanding $a e A_2=B (N n_1-\frac{N}{2}-\frac{N^2}{2}),$
whenever $n_2=N.$ Thus, all the plaquettes have now flux equal to B, 
with the exception of the one starting at $(n_1=N,~~n_2=N).$ This 
plaquette is characterized by flux equal to $-B (N^2-1),$ which is 
exactly what is needed to cancel the contribution of all other plaquettes.
We remind the reader that the value of these plaquettes does not 
depend on $x_3,$ so, considering the three-dimensional configuration,
we have a constant flux through the $N^2-1$ plaquettes and a ``Dirac string"
of flux of opposite sign through the (N,N) plaquette for every $x_3.$

What we would like to study is the $<{\overline \Psi} \Psi>$ condensate,
which will depend on the lattice point in which it will be calculated.
The dependence will be very small if one uses sites deep inside the region 
where the magnetic field is homogeneous. One should expect a strong 
dependence on the site near the location of the string of the large 
negative flux. We have measured the condensate in the middle of the 
lattice. Another concern is the kind of boundary conditions one will
use for the fermions. In the time direction we used antiperiodic 
boundary conditions, conforming to common practice. For the spatial
directions one may use periodic or fixed boundary conditions (the latter 
meaning that the fermion fields vanish on the boundaries). We have used 
both and found out that the results agree for large enough lattices.
Finally we chose to work with the fixed boundary conditions, so that 
the fermion matrix feels the least influence from the big negative flux, 
which is located exactly on the boundary.

\section{Lattice Results}

We start with the treatment of the model in the limit $\beta_G 
\rightarrow \infty,$ that is we consider the case where the statistical 
gauge potential is frozen to zero. In this case, the action simplifies to:
\bea
&~&S = \sum_{n,n^\prime} {\overline \Psi}_n 
Q_{n,n^\prime} \Psi_{n^\prime}
\nn \\
&~&Q_{n,n^\prime} = \delta_{n,n^\prime}-K \sum_\mu [\delta_{n^\prime,n+\hat
\mu} \gamma_\mu e^{i g a A_{n \mu}} - \delta_{n^\prime,n-\hat
\mu} \gamma_\mu e^{-i g a A_{n-\hat \mu, \mu}}],
\eea
where $$a e A_1 = -B n_2,~~a e A_2=\delta_{n_2N} B (N n_1-\frac{N}{2}-
\frac{N^2}{2}),~~A_3=0,$$ as explained above.
We now explain some analytical aspects which may give further insight into
the model and help us to have an
independent check of our results in this case. We note that 
for large enough lattices one may disregard the boundary effects, if 
the condensate is measured away from the regions of inhomogeneity, 
as explained above. Then the gauge field configuration is simplified to:
$$a e A_1 = -B n_2,~~A_2=0,~~A_3=0,$$ and the action is further reduced to 
the form:
\bea
&~& S= \sum_{n,n'} {\overline \Psi}_n Q_{nn'} \Psi _{n'} \nn \\
&~&Q_{nn'}=\delta _{nn'} - K \gamma_1 (\delta _{n',n+{\hat 1}}
e^{-iB n_2} - \delta _{n',n-{\hat 1}}e^{iB n_2}) - \nn \\
&~&K\gamma _2 (\delta _{n', n+{\hat 2}}- \delta _{n',n-{\hat 2}}) - 
K \gamma _3 (\delta _{n', n+{\hat 3}} - \delta _{n',n-{\hat 3}}) 
\label{latticeaction}
\eea
We observe from (\ref{latticeaction}) that in order to take the
continuum limit of the lattice action, one should consider $B$
on the lattice, such that the maximum $B N$ of $B n_2$ should satisfy 
${\rm sin}B N \simeq B N$, which implies that \footnote{We thank
G. Tiktopoulos for a useful discussion on this point.} the magnetic field 
B should be small enough. If we would like a somewhat more quantitative
estimate of the maximum allowed magnetic field, a trivial calculation of 
the cyclotron radius can indicate that for $B \simeq \frac{\pi}{8}$ 
this radius equals two lattice spacings and it decreases for increasing B.
Thus the wave functions are expected to have structure at distances smaller 
than one lattice spacing and the discretization becomes important.
Thus it seems reasonable to stop at this value of B.

A semi-analytical evaluation of the fermion condensate 
$<{\overline \Psi} _{(0 n_2 0)} \Psi _{(0 n_2 0)} >$
is achieved by first taking the Fourier transform of 
$Q_{n_2n'_2}$ for the directions 1 and 3, and then   
inverting $Q_{n_2n'_2}$:
\be
<{\overline \Psi} _{(0 n_2 0)} \Psi _{(0 n_2 0)} > 
= \frac{1}{N N_\tau} \sum_{p_1,p_3} (\frac{1}{Q})_{n_2n_2} 
\ee
\be
Q_{n_2n'_2}=\delta _{n_2n'_2} - 2iK\gamma_1 {\rm sin}(p_1 + B n_2)
\delta _{n_2n'_2}
- 2iK\gamma _3 {\rm sin}p_3 \delta _{n_2 n'_2} - 
K\gamma _2 (\delta _{n'_2,n_2+{\hat 1}}
- \delta _{n'_2, n_2-{\hat 1}}). \label{lattcond}
\ee
The quantities $p_1$ and $p_3$ in the sum take the values $\frac{2 \pi}{N} n_1$
and $\frac{2 \pi}{N_\tau} (n_3+\frac{1}{2})$ respectively.
We remark here that this approach parallels the one of Landau when
he first derived the Landau levels for the motion of a particle in
a homogeneous external magnetic field.
We have checked that the results from the full lattice calculation and the 
ones from the procedure outlined in (\ref{lattcond}) agree perfectly.

In figures 4 and 5 we plot $\Delta <{\overline \Psi}\Psi> \equiv$ 
$ <{\overline \Psi}\Psi>|_B-<{\overline \Psi}\Psi>|_{B=0}$ 
versus the magnetic field B. In this and the following figures
we use the ratio $b$, defined by $b \equiv \frac{eB}{eB|_{max}}.$ 
Note that $eB|_{max}=\pi$ (recall the discussion after equation 
(\ref{lact}) ).

%%%%%%%%%%%%%%%%%%%%%%%%%%%%%%%%%%%%%%%%%%%%%%%%%%%%%%%%%%%%%%%%
\begin{figure}
\centerline{\hbox{\psfig{figure=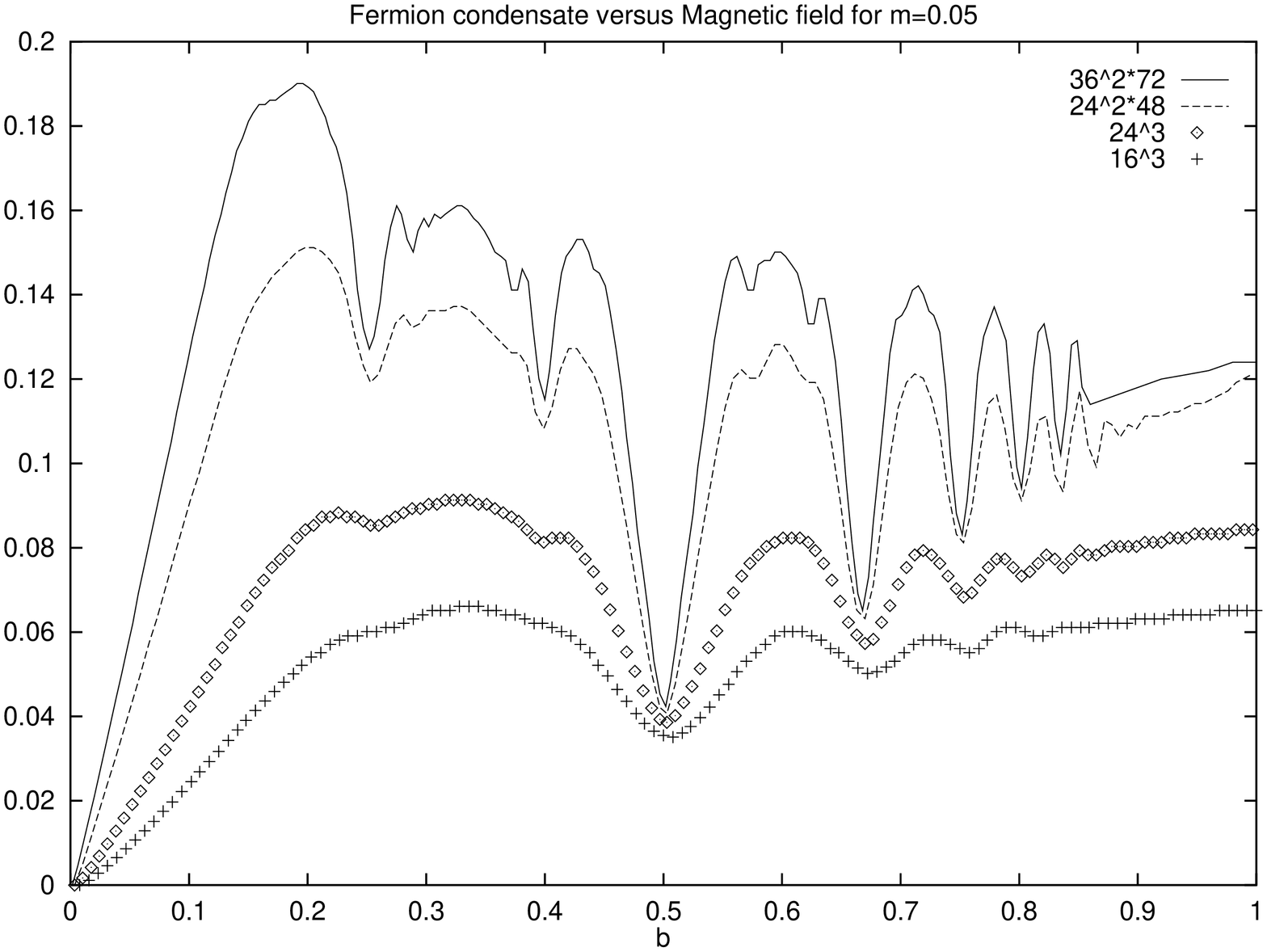,height=9cm,angle=-90}}}
\caption[f4]{$\Delta <{\overline \Psi} \Psi>$ versus  the magnetic field 
for $m=0.05$ and various volumes.}
\label{f4}
\end{figure}
%%%%%%%%%%%%%%%%%%%%%%%%%%%%%%%%%%%%%%%%%%%%%%%%%%%%%%%%%%%%%%%%

Figure 4 shows the behaviour of $\Delta <{\overline \Psi}\Psi>$, when 
the mass parameter $m$ is 0.05 for various lattice volumes.
We see clearly very strong volume effects, as expected for this small 
mass value. As explained in figure 2, the lattice size should be 
around 100, if one is to get strictly zero temperature results. We see
that the condensate keeps increasing with the volume. The finite temperature 
effects are stressed emphatically through the comparison of the results
for the $24^3$ and $24^2 \times 48$ lattices. We should note that 
we expect a non-trivial wave function renormalization of the composite operator $<{\overline \Psi} \Psi>$, which can probably account for the
difference in the values of the condensate as shown in figures 1 and 2 
on one side and figure 4 on the other.
The small B region has qualitatively the same behaviour 
as the corresponding result in figure 1.
One may also observe the complicated 
behaviour for large magnetic fields, which is clearly a lattice 
artifact, as explained previously. 

%%%%%%%%%%%%%%%%%%%%%%%%%%%%%%%%%%%%%%%%%%%%%%%%%%%%%%%%%%%%%%%%
\begin{figure}
\centerline{\hbox{\psfig{figure=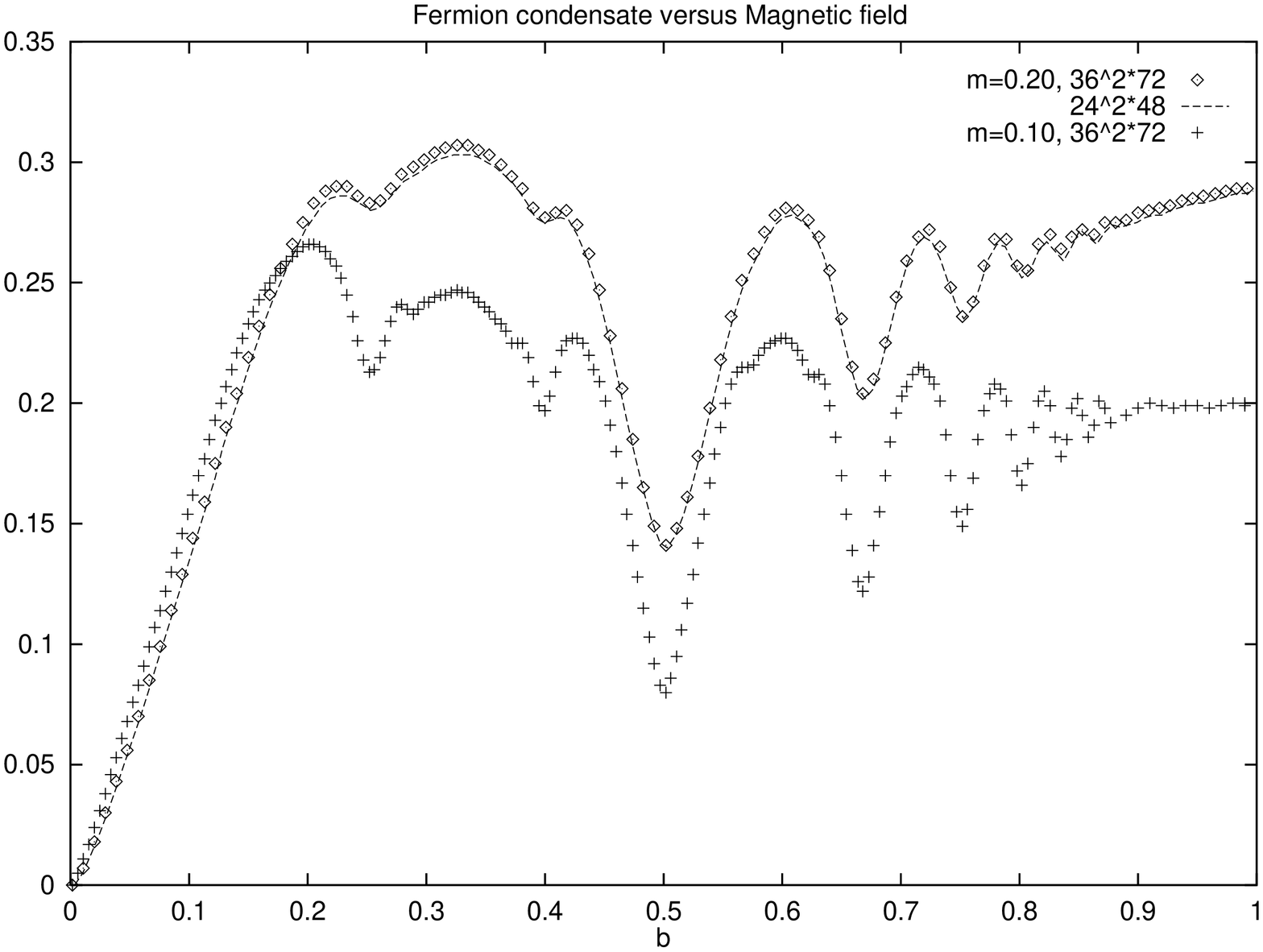,height=9cm,angle=-90}}}
\caption[f5]{$\Delta <{\overline \Psi} \Psi>$ versus  the magnetic field 
for $m=0.10$ and $m=0.20.$}
\label{f5}
\end{figure}
%%%%%%%%%%%%%%%%%%%%%%%%%%%%%%%%%%%%%%%%%%%%%%%%%%%%%%%%%%%%%%%%

Figure 5 depicts the mass dependence of $\Delta <{\overline \Psi}\Psi>:$
Two masses ($m=0.20,$ $0.10$) have been used.
For $m=0.20$ two different volumes have been used ($24^2 \times 48$ 
and $36^2 \times 72$). We recall from the discussion in figure 2 that for
lattices of linear extent 25 or higher we should get zero temperature 
results. This is really what is clearly seen in figure 5: the two lattices,
assumed large enough for this mass, give identical results. 
We have also plotted the results for $m=0.10$ on the same figure.
Two facts have to be pointed out: a) In the small B region we see that the
condensates for $m=0.10$ are bigger than the ones for $m=0.20,$ 
as anticipated in figure 1. b)The condensates go to 
zero with the mass for fixed volume. This just means that 
to get meaningful results one should 
first take the thermodynamic limit and after that the massless limit.

Another issue that has to be settled is gauge invariance, in the sense
that the various gauge potentials that may be used to generate a
homogeneous magnetic
field are equivalent. In figure 6 we show the result for the condensate 
for two gauge choices: $A_1=-B n_2,~A_2=0,~A_3=0,$ and 
$A_1=-\frac{B n_2}{2},~A_2=\frac{B n_1}{2},~A_3=0.$ Lattice 
volumes $6^3$ and $16^3$ have been used. We observe an approximate 
gauge invariance for the small volume and exact gauge invariance 
for the large volume. The discrepancy in the small volume has to do with 
the fact that only approximate gauge invariance can be achieved on small
lattices, which are most sensitive to the boundary conditions.

%%%%%%%%%%%%%%%%%%%%%%%%%%%%%%%%%%%%%%%%%%%%%%%%%%%%%%%%%%%%%%%%
\begin{figure}
\centerline{\hbox{\psfig{figure=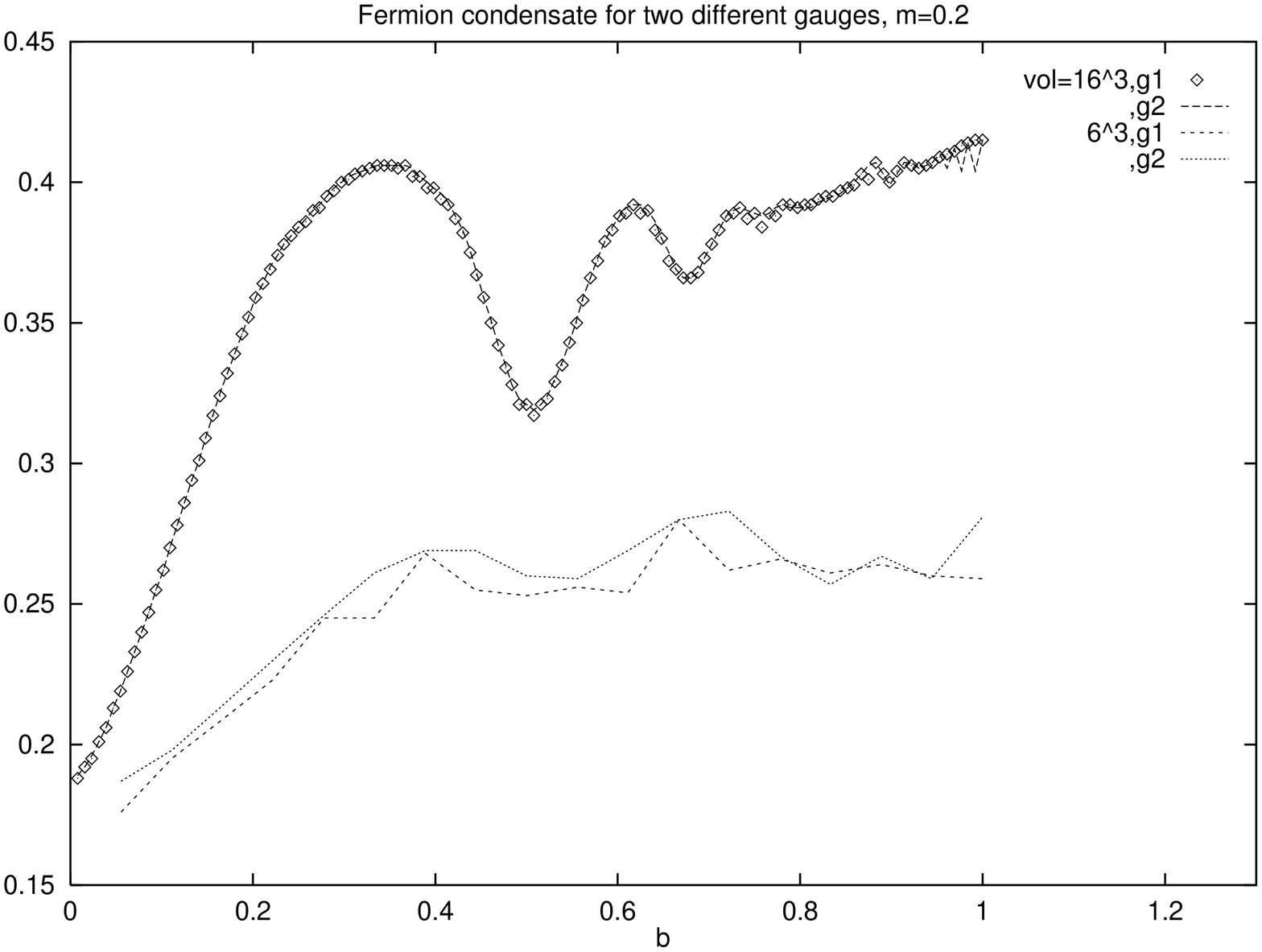,height=9cm,angle=-90}}}
\caption[f6]{$<{\overline \Psi} \Psi>$ versus the magnetic field 
for two gauge choices and two lattice volumes.}
\label{f6}
\end{figure}
%%%%%%%%%%%%%%%%%%%%%%%%%%%%%%%%%%%%%%%%%%%%%%%%%%%%%%%%%%%%%%%%

After the discussion on the free fermion case, we next proceed to
incorporate quantum fluctuations of the gauge field in the analysis.

%%%%%%%%%%%%%%%%%%%%%%%%%%%%%%%%%%%%%%%%%%%%%%%%%%%%%%%%%%%%%%%%
\begin{figure}
\centerline{\hbox{\psfig{figure=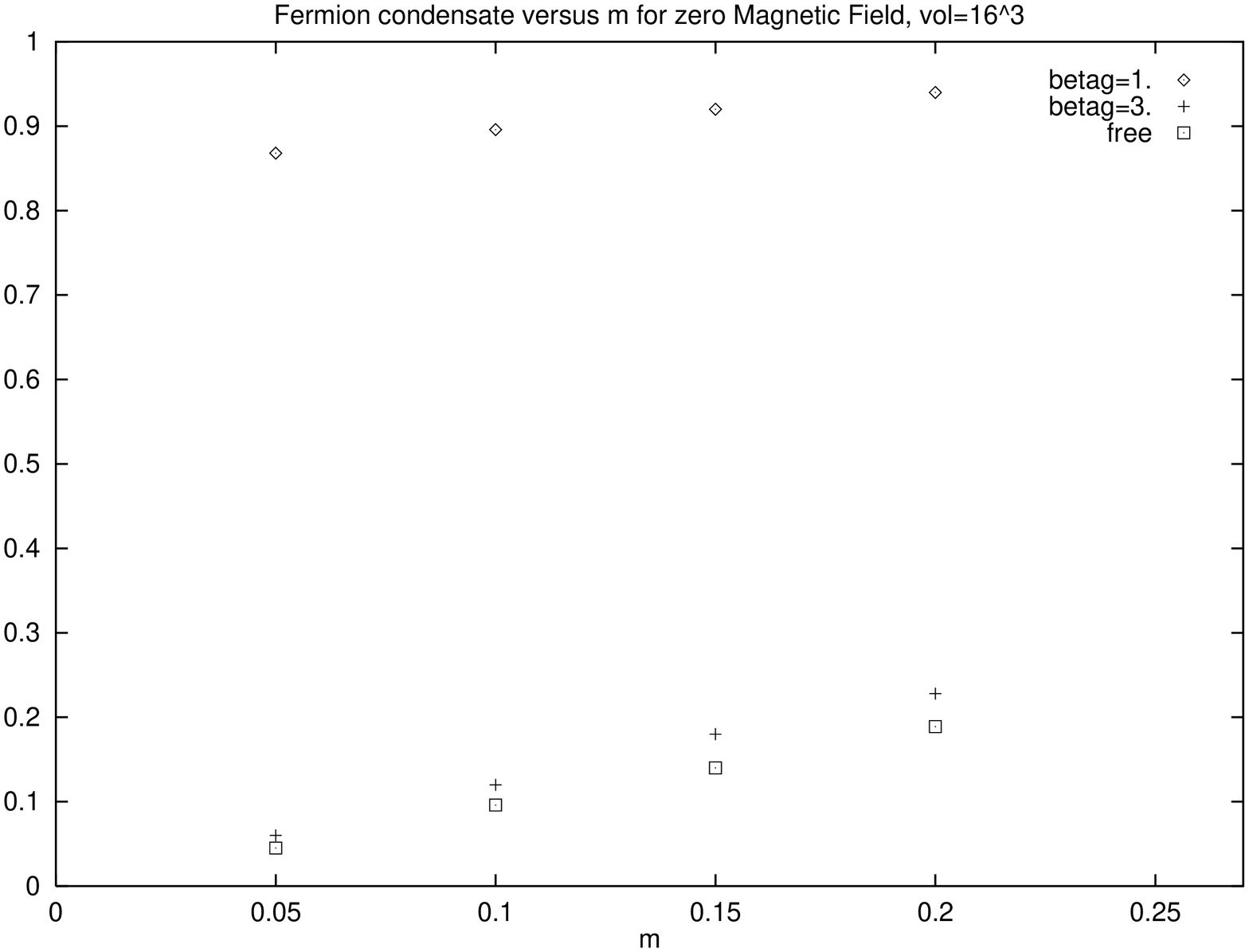,height=9cm,angle=-90}}}
\caption[f7]{$<{\overline \Psi} \Psi>$ versus m for zero magnetic field 
and three values of the gauge coupling.}
\label{f7}
\end{figure}
%%%%%%%%%%%%%%%%%%%%%%%%%%%%%%%%%%%%%%%%%%%%%%%%%%%%%%%%%%%%%%%%

As a first step, we show in figure 7 the condensate for zero magnetic 
field versus the mass 
$m$ for a $16^3$ lattice and three values of the gauge coupling
$\beta_G:$ 1, 3 and infinite (the latter corresponds to the free field).
The value 1 is in the strong coupling regime, where a condensate 
will be formed dynamically, while the value 3 is in the 
weak coupling region, where 
the condensate is small (could be zero if we had dynamical flavors with more
than four flavours). We observe that the values for $\beta_G=3$ are
very close to the ones for the free field theory; the extrapolation to 
zero mass for $\beta_G=3$ will yield a very small value. On the other hand, 
the condensate for $\beta_G=1$ are impressively big, signalling the 
breakdown of the global $SU(2)$ symmetry; 
the extrapolation to the massless limit certainly points to a non-zero value.
We should point out here that we should subtract the perturbative 
contribution from our lattice result. In that sense our result is 
preliminary, since it lacks the perturbative calculation. 
We plan to include this in a future publication.

In the last two figures we depict the results for the full model versus 
the magnetic field. 
In figure 8 one may see the $16^3$ result for $\beta_G=3,~m=0.20$
These results should be compared against the  ones in figure 6 
for $V=16^3,~\beta_G=3,~m=0.20.$ They are obviously very similar,
as expected.

%%%%%%%%%%%%%%%%%%%%%%%%%%%%%%%%%%%%%%%%%%%%%%%%%%%%%%%%%%%%%%%%
\begin{figure}
\centerline{\hbox{\psfig{figure=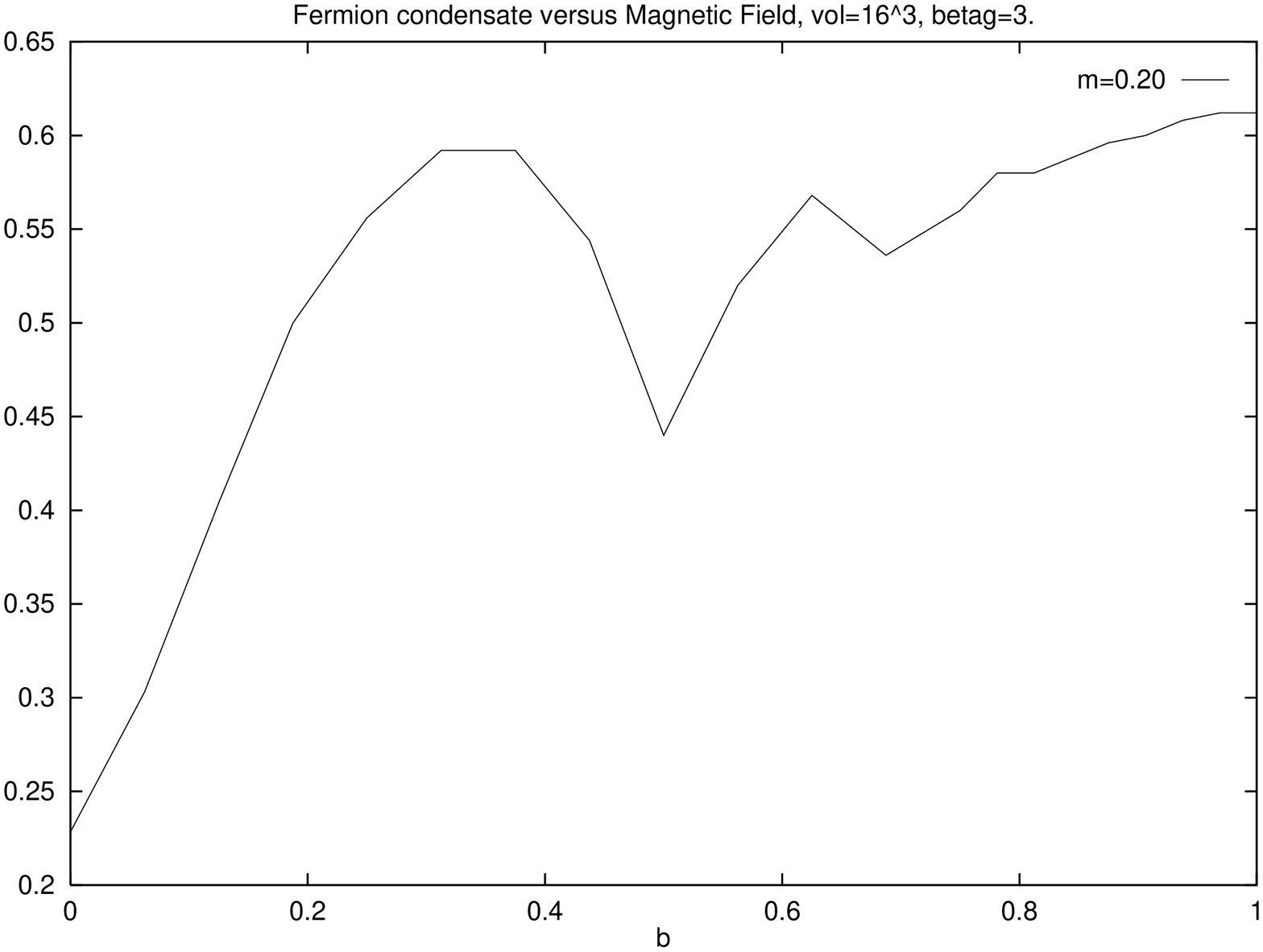,height=9cm,angle=-90}}}
\caption[f8]{$<{\overline \Psi} \Psi>$ versus the magnetic field 
for $m=0.20$ and $\beta_G=3,$ for a $16^3$ lattice.}
\label{f8}
\end{figure}
%%%%%%%%%%%%%%%%%%%%%%%%%%%%%%%%%%%%%%%%%%%%%%%%%%%%%%%%%%%%%%%%

In figure 9 we show the results for $16^3$ 
lattice, $\beta_G=1$ and $m=0.20,~0.10,~0.05.$ We observe the same 
qualitative behaviour with the free case in the small B region. 
However there is a crucial difference: a non-zero condensate is formed,
{\em however small the bare mass may be,} because of the gauge interaction.
On the other hand, we find an impressive smoothening out of the
graph in the large B region; moreover, the difference between the 
results for the various masses decrease with increasing B, so, taken
at face value, these results imply mass independence of the results in the 
large B region, as expected on the basis of the continuum results.
However, in this region we expect many lattice
artifacts, as explained before, so one should be careful with 
the interpretation of the subtle continuum limit.

%%%%%%%%%%%%%%%%%%%%%%%%%%%%%%%%%%%%%%%%%%%%%%%%%%%%%%%%%%%%%%%%
\begin{figure}
\centerline{\hbox{\psfig{figure=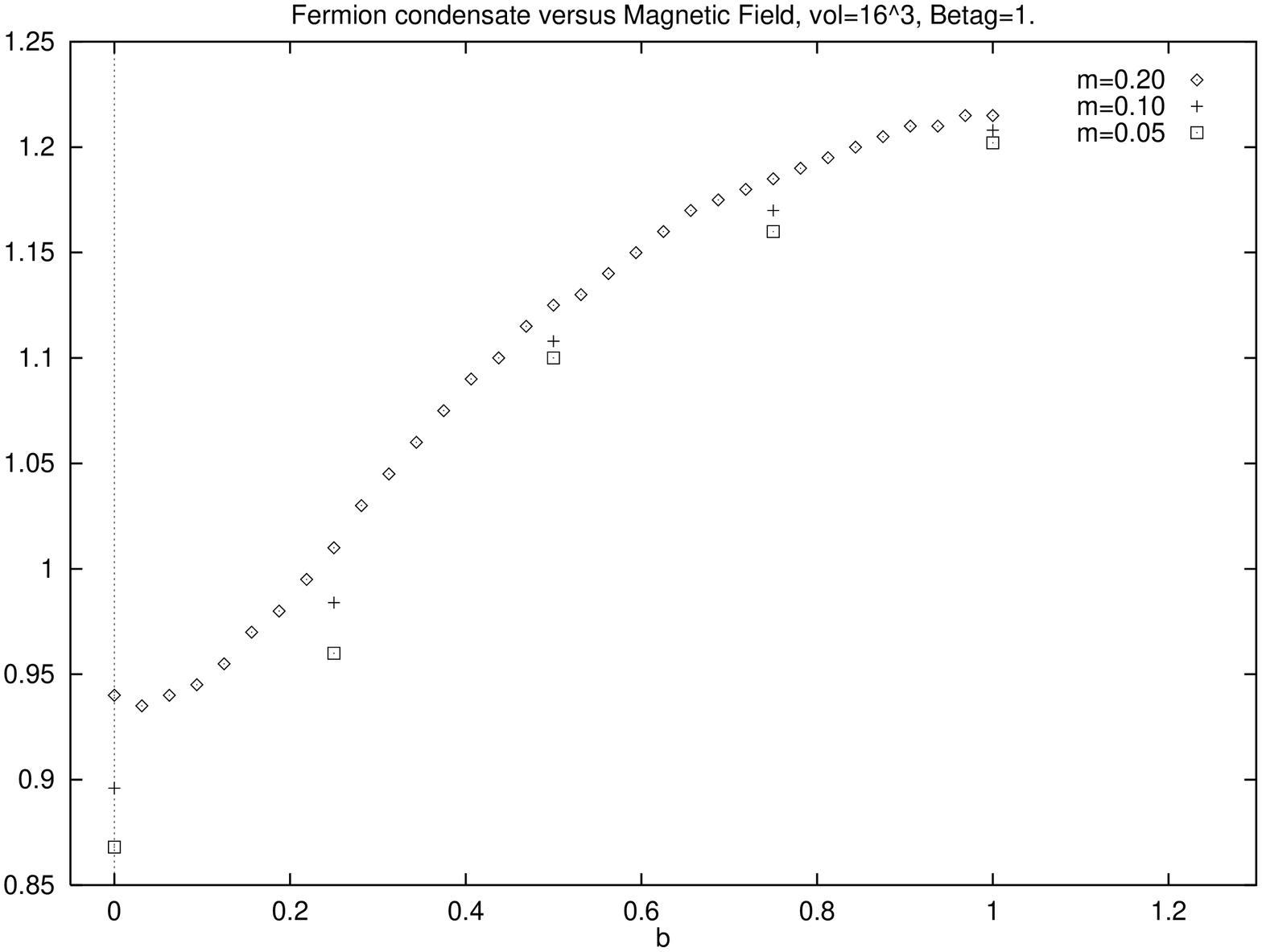,height=9cm,angle=-90}}}
\caption[f9]{$<{\overline \Psi} \Psi>$ versus m for zero magnetic field 
for $m=0.20,~m=0.10,~m=0.05$ and $\beta_G=1,$ for a $16^3$ lattice.}
\label{f9}
\end{figure}
%%%%%%%%%%%%%%%%%%%%%%%%%%%%%%%%%%%%%%%%%%%%%%%%%%%%%%%%%%%%%%%%

\newpage
\paragraph{}
\noindent {\Large {\bf Acknowledgements}}
\paragraph{}

The authors wish to thank G. Tiktopoulos for valuable discussions.
G.K. wishes to acknowledge partial financial support 
from PENED 95 Program No. 1170 of the Greek General Secretariat
of Research and Technology. K.F. wishes to thank the TMR project 
number FMRX-CT97-0122 for partial financial support.

\end{document}